\documentstyle[epsfig]{aipproc}

\begin{document}

\title{FINITE-DIFFERENCE CALCULATIONS FOR ATOMS AND DIATOMIC MOLECULES
IN STRONG MAGNETIC AND STATIC ELECTRIC FIELDS}

\author{Mikhail V. Ivanov$^\dag$ and Peter Schmelcher}
\address{Theoretische Chemie, Physikalisch--Chemisches Institut,\\
Universit\"at Heidelberg, INF 229, D-69120 Heidelberg,\\
Federal Republic of Germany\\
e-mail: Mikhail.Ivanov@tc.pci.uni-heidelberg.de\\
\dag Permanent address: Institute of Precambrian Geology and Geochronology,\\
Russian Academy of Sciences,\\
Nab. Makarova 2, St. Petersburg 199034, Russia\\
e-mail: MIvanov@MI1596.spb.edu
}

\maketitle

\begin{abstract}
Fully numerical mesh solutions of 2D quantum equations of 
Schr\"odinger and Hartree-Fock type allow us to work with 
wavefunctions which possess a very flexible geometry.
This flexibility is especially important
for calculations of atoms and molecules in strong external fields where neither the external
field nor the internal interactions can be considered as a perturbation. The applications of the
present approach include calculations of atoms and diatomic molecules in strong static
electric and magnetic fields. For the latter we
have carried out Hartree-Fock calculations for
He, Li, C and several other atoms. 
This yields in particular the first
comprehensive investigation of the ground state configurations of the Li and C atoms
in the whole range of magnetic fields ($0<B<10000$ a.u.) and a study of the ground state
electronic configurations of all the atoms with $1<Z\leq 10$
and their ions ${\rm A^+}$ in the high-field fully
spin-polarised regime. 
The results in a case of a strong electric field relate to single-electron
systems including the correct solution of the Schr\"odinger equation 
for the ${\rm H_2^+}$ ion (energies
and decay rates) and the hydrogen atom in strong 
parallel electric and magnetic fields.
\end{abstract}

\section{Introduction}

Theoretical studies of atoms and molecules
in strong external fields are 
motivated by several applications. 
The latter are e.g. 
experiments with intense laser
beams (electromagnetic fields with dominating electric component) 
and astronomical observations of white dwarfs and neutron stars 
(magnetic fields).
The experimental availability of extremely strong electric fields in laser
beams makes the theoretical study of various
atomic and molecular species under such conditions 
very desirable.
The properties of atomic and molecular systems in
strong fields undergo dramatic changes in comparison with
the field-free case.
These changes are associated with the strong distortions of
the spatial distributions of the electronic density and correspondingly
the geometry of the electronic wavefunctions.
This complex geometry is difficult for its description by means
of traditional sets of basis functions and requires
more flexible approaches which can, in particular, 
be provided by multi-dimensional mesh
finite-difference methods.

Let us discuss the problem of atoms in a strong magnetic field
in more detail.
We start this consideration with the hydrogen atom
which was the first atom whose
behaviour in strong magnetic fields was investigated
(for a list of references see \cite{Fri89,RWHR,Ivanov88,Kra96}).
In cylindrical coordinates $(\rho,z)$ its non-relativistic Hamiltonian has the form
(we use atomic units throughout our work)
\begin{eqnarray}
H=-\frac 12\left(\frac{\partial ^2}{\partial \rho ^2}+
\frac 1\rho \frac \partial
{\partial \rho }+\frac{\partial ^2}{\partial z^2}-
\frac{m^2}{\rho ^2}\right)
+s_z\gamma+\frac{m}{2}\gamma+\frac{\gamma^2}{8}\rho^2
-\frac 1r
\label{eq:HFSCHam}
\end{eqnarray}
where $m$ is the magnetic quantum number, $s_z$ is the spin $z$ projection
and $\gamma=B/B_0$, $B_0=\hbar c/ea_0^2=2.3505 {\cdot} 10^5$T is the
magnetic field strength in atomic units.
The magnetic field is parallel to the $z$ axis.
The Hamiltonian (\ref{eq:HFSCHam}) contains two potentials of different
spatial symmetries: the spherical-symmetric Coulomb term $1/r$ and
the cylindrically symmetric potential of the magnetic field
$\gamma^2 \rho^2 /8$.
When considering the impact of the competing Coulomb and diamagnetic interaction
it is reasonable to distinguish the three different regimes of weak, strong and
intermediate fields.
In the latter case the magnetic and Coulomb forces are comparable.
In the case of relatively weak fields the main features of
the geometry of the wavefunction are determined by the dominating
Coulomb term whereas the effect of the magnetic field can be consider as
a perturbation of the Coulomb wavefunctions.
For the opposite situation of very strong magnetic fields
and dominating cylindrical symmetry the adiabatic
approximation \cite{ElLoudon,Neuhauser,Godefroid} 
was the main theoretical tool during the last four decades.
This approximation separately considers the fast motion of the electron across the field
and its slow motion in a modified Coulomb potential along the field direction.
Both early (see \cite{Garstang}) and more recent works
\cite{RWHR,SimVir78,Friedrich,Fonte,Schmidt}
on the hydrogen atom have used different approaches for these regimes
of the magnetic field.
All these calculations had problems when considering the hydrogen
atom in fields of intermediate strength.
The detailed calculations of the hydrogen energy levels carried out
by R\"osner {\it et al} \cite{RWHR} also retained the separation into
different regimes of the field strength by decomposing
the electronic wave function either in terms of spherical
(weak to intermediate fields) or cylindrical (intermediate to high fields) orbitals.
A solution allowing to obtain comprehensive results 
on low-lying energy levels of the
hydrogen atom for arbitrary field strengths 
including the intermediate field regime is provided by
the multi-dimensional mesh solution 
of the Schr\"odinger equation \cite{Ivanov88}.

For different electronic degrees of excitation of the atom the intermediate
regime is met for different absolute values of the field strength.
For the ground state this regime is roughly given by $\gamma=~0.2-20$. 
For atoms with several electrons there are two decisive factors which
enrich the possible changes in the electronic structure with varying
field strength compared to the one-electron system.
First, we have a third competing interaction which is
the electron-electron repulsion and, second, the different electrons
feel very different Coulomb forces, i.e. possess different one particle energies,
and consequently the regime of the intermediate field strengths
appears to be the sum of the intermediate regimes for the separate electrons.
  
The fact that most methods have problems in the intermediate field region
has consequences for the current state of the art of our knowledge on 
on multi-electron atoms in strong magnetic fields.
There exist a number of such investigations in the literature
\cite{Neuhauser,Godefroid,Mueller75,Virtamo76,Larsen,Gadiyak,VinBay89,Ivanov91,TKBHRW,Ivanov94,JonesOrtiz,JonesOrtiz97}.
The majority of them deals with the adiabatic regime in superstrong fields and
the early works are mostly Hartree-Fock (HF) type calculations.
There are also several early variational calculations for the low-field domain
\cite{Larsen,Henry74,Surmelian74}. HF calculations for arbitrary field strengths
have been carried out in refs.\cite{RWHR,TKBHRW} by applying
two different sets of basis functions in the high- and low-field regimes. As a result
of the complicated geometry this approach suffers in the intermediate regime
from very slow convergence and low accuracy of the calculated energy eigenvalues.
Accurate HF calculations for arbitrary field strengths were carried out in refs.
\cite{Ivanov91,Ivanov94} by the 2D mesh HF method. Investigations on the
ground state as well as a number of excited states of helium including
the correlation energy have recently been performed via a Quantum Monte
Carlo approach \cite{JonesOrtiz97}. 
Very recently benchmark results with
a precision of $10^{-4}-10^{-6}$ for the energy levels
have been obtained for a large number of
excited states with different symmetries using a configuration
interaction approach with an anisotropic Gaussian basis set \cite{Bec99,Bec2000}.
Focusing on systems with more than two electrons however the number of investigations is very 
scarce \cite{Neuhauser,JonesOrtiz,Muell84}.
In view of the above there is a need for further quantum mechanical
investigations and for data on atoms with more than two electrons in a strong magnetic field.
For the carbon atom there exist two investigations \cite{Neuhauser,Godefroid} 
in the adiabatic approximation
which give a few values
for the binding energies in the high field regime and one more relevant recent
work by Jones {\it et al} \cite{JonesOrtiz}.
On the other hand, our two-dimensional Hartree-Fock approach allowed us
recently to perform precise and reliable consideration of a series of
multi-electron atoms for the whole range of the magnetic field
strengths from $\gamma=0$ up to ultrastrong fields $\gamma=10^3-10^4$ 
\cite{Ivanov91,Ivanov94,Ivanov98,IvaSchm98,IvaSchm99,IvaSchm2000}.

\section{Two-dimensional mesh Hartree-Fock method}

Our calculations for multi-electron atoms in magnetic fields are carried out
under the assumption of an infinitely heavy nucleus
in the (unrestricted) Hartree-Fock approximation.
The solution is established in the cylindrical coordinate system
$(\rho,\phi,z)$ with the $z$-axis oriented along the magnetic field.
We prescribe to each electron a definite value of the magnetic
quantum number $m_\mu$.
Each single-electron wave function $\Psi_\mu$ depends on the variables
$\phi$ and $(\rho,z)$
\begin{eqnarray}
\Psi_\mu(\rho,\phi,z)=(2\pi)^{-1/2}e^{-i m_\mu\phi}\psi_\mu(z,\rho)
\label{eq:phiout}
\end{eqnarray}
where $\mu$ denotes the numbering of the electrons.
The resulting partial differential equations for $\psi_\mu(z,\rho)$
have been presented in
ref.\cite{Ivanov94}.

The one-particle equations for the wave functions $\psi_\mu(z,\rho)$
are solved
by means of the fully numerical mesh method described in refs.
\cite{Ivanov88,Ivanov91,Ivanov94}.
In our first works on the helium atom in magnetic fields
\cite{Ivanov91,Ivanov94} we calculated
the Coulomb and exchange integrals by means
of a direct summation over the mesh nodes.
But this direct method is very expensive with respect to 
the computer time and due to this reason we obtained in the following
works \cite{Ivanov98,IvaSchm98,IvaSchm99,IvaSchm2000}
these potentials as solutions
of the corresponding Poisson equation.
The problem of the boundary conditions for the Poisson equation
as well as
the problem of simultaneously solving
Poisson equation on the same meshes with
Schr\"odinger-like equations for the wave functions $\psi_\mu(z,\rho)$
have been discussed in ref.\cite{Ivanov94}.

The simultaneous solution of the Poisson equations for the
Coulomb and exchange potentials and
Schr\"odinger-like equations for the wave functions $\psi_\mu(z,\rho)$
is a complicated computational problem, especially
for atoms in strong magnetic fields.
The problem consists in the different geometry of 
the spatial distribution of the electron density and
the potentials correspondent to this density.
In strong magnetic fields the distributions of the electronic densities
are compressed towards the $z$ axis and look like needles directed
along the $z$ axis.
The equations for the wavefunctions can be solved in finite cylindrical
domains as done in refs. \cite{Ivanov88,Ivanov91,Ivanov94}.
For strong magnetic fields $\gamma>>1$ these domains can be rather small
in the $\rho$ direction.
On the other hand, the potentials created by these charge distributions
cannot have such a strongly anisotropic form and the Poisson equations
for them must be solved on meshes with the distribution of nodes
not very different for the $z$ and $\rho$ directions.
This means some loss of the precision for the wavefunctions due to
a decrease of the number of nodes in the area of a large electronic density.
The most difficult problem, however, is the different asymptotic behaviour of
the wavefunctions and potentials. 
The wavefunctions of the bound electrons decrease exponentially
as $r\rightarrow \infty$ ($r$ is the distance from the origin).
This simplifies the problem of the solution of the corresponding
equations in the infinite space because it is possible either to solve
these equations in a finite domain $\Omega$
(with simple boundary conditions $\left. \psi_{_{}} \right|_{\partial\Omega}=0$
or $\left. \partial \psi_{_{}}/\partial n \right|_{\partial\Omega}=0$)
with negligible errors for
domains of reasonable dimensions or otherwise 
to solve these equations in the infinite
space on meshes with exponentially growing distances
between nodes as $r\rightarrow \infty$.
The solutions of the Poisson equations for non-zero sums of charges
decrease as $1/r$ as $r\rightarrow \infty$.
In result, every spatial restriction of the domain $\Omega$
introduces a significant error into the final solution.
In some other mesh Hartree-Fock approaches developed for diatomic molecules
(e.g. see \cite{LaaPyy84,Kobus93}) this problem
has been solved for finite $\Omega$ by introducing special
boundary conditions for the potentials obtained from
the asymptotic behaviour of the potentials.
This approach, being in principle approximate, requires
additional calculations with different extensions of $\Omega$
to estimate the error.

In the present approach we address the above problems 
by using special forms of non-uniform meshes \cite{Ivanov98}.
Solutions to the Poisson equation on separate meshes contain
some errors $\delta_P$ associated with an inaccurate description of the
potential far from the nucleus.
However due to the special form of the function $\delta_P(h)$
for these meshes
(where $h$ is a formal mesh step)
the errors do not show up in the final results for the energy
and other physical quantities, which we obtain by means of the Richardson
extrapolation procedure (polynomial extrapolation to $h=0$
\cite{Ivanov88,ZhVychMat}).
The main requirement for these meshes is not an exponential, but
a polynomial increase of the mesh step $h$ when $r\rightarrow \infty$.
Moreover, this behaviour can be only linear one,
i.e. $h^{-1}=O(1/r)$ as $r\rightarrow \infty$.
The error of the mesh solution in this case has the form
of a polynomial of the formal step of the mesh $\tilde{h}=1/N$,
where $N$ is the number of nodes along one of the coordinates.
In practical calculations these meshes are introduced by means
of an orthogonal coordinate transformation from
the physical coordinates $x_{\rm p}$
to the mathematical ones $x_{\rm m}$ made separately
for $\rho$ and $z$.
And the numerical solution is, in fact, carried out
on uniform meshes in the mathematical coordinates $x_{\rm m}$.
The characteristic feature of these meshes consists of rapidly increasing 
coordinates of several outermost nodes when increasing the total number
of nodes and decreasing the actual mesh step in the vicinity of the origin.
Due to this property we call these meshes
``Run away'' meshes.
To be concrete we present here two such meshes:

1. The ``Plain Poisson'' mesh is generated by the coordinate transformation 
\begin{eqnarray}
x_{\rm p}=A\frac{x_{\rm m}}{1-x_{\rm m}^2}
\label{eq:PlainPoiss}
\end{eqnarray}
$-\infty<x_{\rm p}<+\infty$, $-1<x_{\rm m}<+1$, $A$ is a constant.
This simplest mesh of this group is near to the uniform ones
near the origin (i.e. a plot of the distance between the neighbouring 
nodes contains a large horizontal section close to $x_{\rm p}=0$)
and then this mesh smoothly transforms to the ``run away''
behaviour for $x_{\rm p}\rightarrow \infty$.

2. The ``Atomic Poisson'' mesh
\begin{eqnarray}
x_{\rm p}=A\frac{(|x_{\rm m}|+b)x_{\rm m}}{1-x_{\rm m}^2}
\label{eq:AtPoiss}
\end{eqnarray}
($b>0$) allows obtaining more precise results for atoms
at reasonable values of $b<1$ due to a more dense distribution of
nodes near the origin.
In fact, this formula provides three different types of behaviour 
in three different domains: 
(a) A uniform mesh in a small vicinity of $x_{\rm p}=0$. 
This behaviour provides absence of irregularities in the finite-difference 
representation of the Hamiltonian. 
(b) $|x_{\rm p}|\approx A|x_{\rm m}^2|$ - the quadratic expansion of the mesh. 
(c) $h^{-1}=O(1/r)$ as $r\rightarrow \infty$.  
The distribution of nodes for not too big distances from the origin
(b) given by the simple formula (\ref{eq:AtPoiss})
is similar to well known ``Lagrange meshes'' for atoms 
and provide
similar precision of results.
(See e.g. \cite{VMB93R} for a definition 
of the ``Laguerre mesh'' 
(a mesh with nodes at zeros of the Laguerre polynomials) 
as a ``Lagrange mesh'' suitable for systems with the Coulomb potential 
and \cite{AbramSteg} (22.16.8) for approximate formulas 
for the zeros of the Laguerre polynomials).

The overall precision of our results depends, of course, on the number of mesh nodes
and, if necessary, can be improved in calculations with denser meshes.
The most dense meshes which we could use in the present calculations
had $120 \times 120$ nodes.
In most cases Richardson's sequences of
meshes with maximal number $80\times 80$ or $60\times 60$ were
sufficient.

\section{The structure of the atomic ground state configurations for
the limit $\gamma\rightarrow\infty$}
\vspace*{-0.7cm}

In this section we provide some qualitative considerations
on the problem of the ground states of multi-electron atoms
in the high field limit.
These considerations
along with the well known electronic structure of
the ground states at $\gamma=0$
present a starting point for the combined
qualitative and numerical considerations given in the following section.
At very high field strengths the nuclear attraction energies and
HF potentials (which determine the motion along the $z$ axis)
are small compared to the interaction energies
with the magnetic field
(which determines the motion perpendicular to the magnetic field
and is responsible for the Landau zone structure of the spectrum).
Thus in the limit ($\gamma \rightarrow \infty $),
all the one-electron wave functions of the ground state belong
to the
lowest Landau zones, i.e. $m_\mu\leq 0$
for all the electrons, and the system must be fully spin-polarised,
i.e. $s_{z\mu}= -{1\over2}$.
For the Coulomb central field the one electron levels form
quasi 1D Coulomb series with the binding energy
$E_B={1\over{2n_z^2}}$ for $n_z>0$,
whereas $E_B(\gamma \rightarrow \infty)\rightarrow \infty$ for $n_z=0$,
where $n_z$ is the number of nodal
surfaces of the wave function crossing the $z$ axis.
In the limit $\gamma \rightarrow \infty$ the ground state wave function
must be formed of the tightly bound single-electron functions with $n_z=0$.
The one-particle binding energies of these functions
decrease as $|m|$ increases and,
thus, the electrons must occupy orbitals with increasing $|m|$ starting
with $m=0$.

In the language of the Hartree-Fock approximation
the ground state wave function of an atom in the high-field limit
is a fully spin-polarised set of single-electron orbitals
with no nodal surfaces crossing the $z$ axis and with
non-positive magnetic quantum numbers decreasing
from $m=0$ to $m=-N+1$, where $N$ is the number of electrons.
In result, we have for the first 10 atoms and positive ions
in the limit $\gamma\rightarrow\infty$
the following structure of ground state configurations which is
a simple substitute of the periodic law at very strong magnetic fields

\vspace*{0.3cm}

\begin{tabular}{@{}llllllllll}
H  &${\rm He^+}$&$1s$&$M=0$&$S_z=-1/2$\\

He \  &${\rm Li^+}$&$1s2p_{-1}$&$M=-1$&$S_z=-1$\\

Li &${\rm Be^+}$&$1s2p_{-1}3d_{-2}$&$M=-3$&$S_z=-3/2$\\

Be &${\rm B^+}$&$1s2p_{-1}3d_{-2}4f_{-3}$&$M=-6$&$S_z=-2$\\

B  &${\rm C^+}$&$1s2p_{-1}3d_{-2}4f_{-3}5g_{-4}$&$M=-10$&$S_z=-5/2$\\

C  &${\rm N^+}$&$1s2p_{-1}3d_{-2}4f_{-3}5g_{-4}6h_{-5}$&$M=-15$&$S_z=-3$\\

N  &${\rm O^+}$&$1s2p_{-1}3d_{-2}4f_{-3}5g_{-4}6h_{-5}7i_{-6}$&$M=-21$&$S_z=-7/2$\\

O  &${\rm F^+}$&$1s2p_{-1}3d_{-2}4f_{-3}5g_{-4}6h_{-5}7i_{-6}8j_{-7}$&$M=-28$&$S_z=-4$\\

F  &${\rm Ne^+}$&$1s2p_{-1}3d_{-2}4f_{-3}5g_{-4}6h_{-5}7i_{-6}8j_{-7}9k_{-8}$&$M=-36$&$S_z=-9/2$\\

Ne &${\rm Na^+}$&$1s2p_{-1}3d_{-2}4f_{-3}5g_{-4}6h_{-5}7i_{-6}8j_{-7}9k_{-8}10l_{-9}$&$M=-45$&$S_z=-5$\\

\end{tabular}

We shall often refer in the following
to these ground state configurations
in the high-field limit as $\left|0_N\right>$.
The states $\left|0_N\right>$ possess the complete spin polarisation $S_z=-N/2$.
Decreasing the magnetic field strength,
we can encounter a series of crossovers of the ground state configuration
associated with transitions of one or several electrons
from orbitals with the maximal values for $|m|$ to
other orbitals with a different spatial geometry of the wave function
but the same spin polarisation.
This means the first few crossovers can take place within the space of fully
spin polarised configurations.
We shall refer to these configurations by noting only the difference
with respect to the state $\left|0_N\right>$.
This notation can, of course, also be extended to non-fully spin
polarised configurations.
For instance the state $1s^22p_{-1}3d_{-2}4f_{-3}5g_{-4}$ with $S_z=-2$ of
the carbon atom can be briefly referred to as $\left|1s^2\right>$,
since the default is the occupation of the hydrogenic series
$1s, 2p_{-1}, 3d_{-2},\ldots$ and
only deviations from it are recorded by this notation.

\section{Ground state electronic configurations
of atoms and positive ions at arbitrary field strengths}

Currently the carbon atom is the most complicated system with
a thoroughly investigated structure of its electronic configurations
for arbitrary magnetic fields and 
we start this section with a consideration of this atom.

In the case of decreasing the magnetic field strength from very large
values to $\gamma=0$ the fully spin-polarised ground state configuration 
of a multi-electron atom must undergo 
one or several crossovers to become finally
the zero-field ground state configuration.
This configuration for the carbon atom  corresponds
to the spectroscopic term $^3P$.
In the framework of the non-relativistic consideration
this term consists of nine states
degenerate due to three possible $z$-projections of the total spin $S_z=-1,0,1$
and three possible values of the total magnetic quantum number $M=-1,0,1$.
For very weak magnetic fields it is reasonable to expect values
$S_z=-1$ and $M=-1$ for the ground state which can be
described in our notation as $1s^2 2s^2 2p_0 2p_{-1}$.

Thus the possible ground state configurations of the carbon atom
can be divided into three groups
according to their total spin projection $S_z$ :
the $S_z=-1$ group (low-field ground state configurations),
the intermediate group $S_z=-2$ and the
$S_z=-3$ group (the high-field ground state configurations).
This grouping is required for the qualitative part of
the following considerations which are based on the geometry
of the spatial parts of the one electron wave functions.

\begin{figure}[b!] % fig 1
\centerline{\epsfig{file=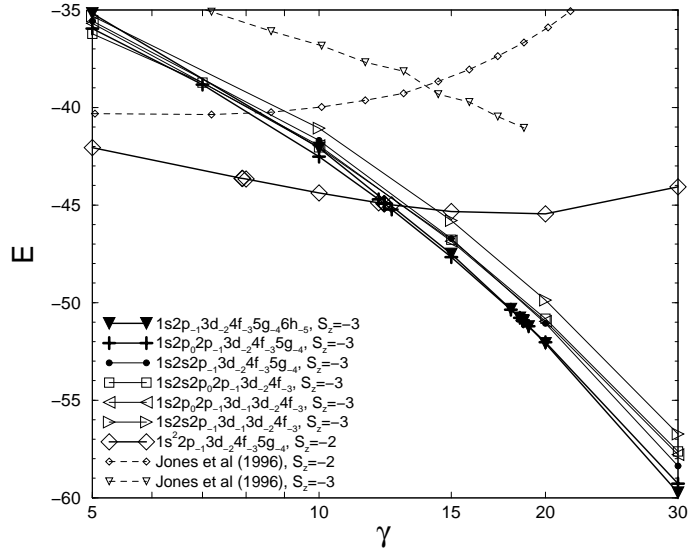,angle=270,width=9cm}}
\vspace{10pt}
\caption{The total energies (in atomic units) of the states
of the carbon atom as functions of the magnetic field strength
considered for the extraction of the ground state electronic configurations
with $S_z=-3$. Our results (solid lines) and data taken from ref.\ref{one}
 (broken lines).
Energies and field strengths are given in atomic units.}
\label{foobar:c_f4}
\end{figure}

\begin{figure}[b!]
\centerline{\epsfig{file=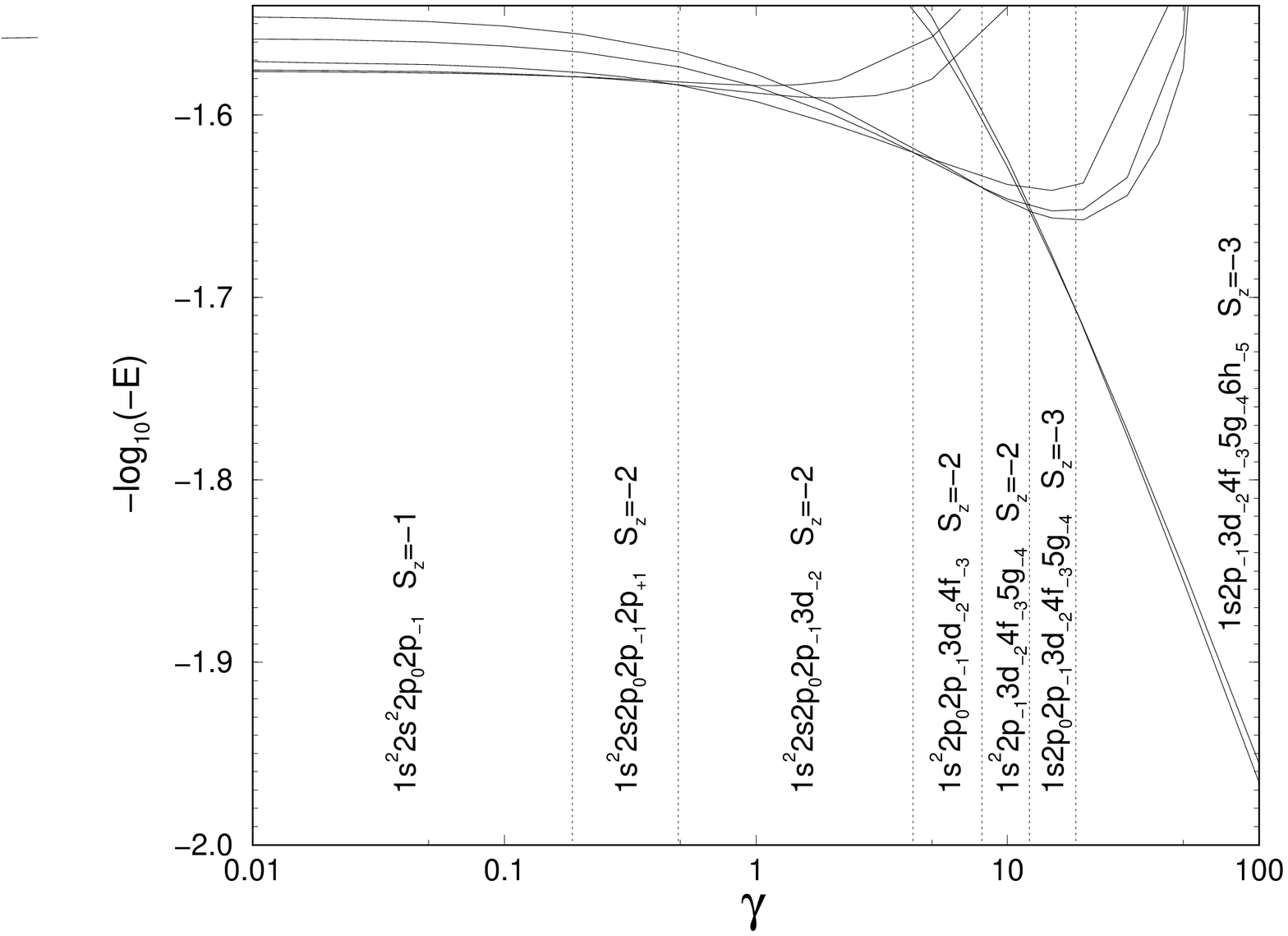,angle=270,width=9cm}}
\vspace{10pt}
\caption{Energies of the ground state configurations
as a function of the field strength. Vertical dotted lines
divide regions belonging to different Hartree-Fock
ground state configurations.}
\label{foobar:c_f5}
\end{figure}

We start our consideration for $\gamma\ne0$  with the high-field ground state and
subsequently consider other possible candidates in question
for the electronic ground state for $S_z=-3$
(see Figure~\ref{foobar:c_f4}) {\it{with decreasing field strength}}.
All the one electron wave functions of the high-field ground state
$1s2p_{-1}3d_{-2}4f_{-3}5g_{-4}6h_{-5}$
possess no nodal surfaces crossing the $z$-axis and occupy the energetically lowest orbitals
with magnetic quantum numbers ranging from $m=0$ down to $m=-5$.
We shall refer to the number of the nodal surfaces crossing the $z$ axis
as $n_z$. The $6h_{-5}$ orbital possesses the smallest binding energy of all orbitals constituting
the high-field ground state. Its binding energy decreases rapidly with decreasing
field strength.  Thus, we can expect that the first crossover of ground state configurations
happens due to a change of the $6h_{-5}$ orbital into one
possessing a higher binding energy at the corresponding lowered range of field strength.
It is natural to suppose that the first transition while decreasing the magnetic field strength
will involve a transition from an orbital possessing $n_z=0$ to one for $n_z=1$.
The energetically lowest available one particle state with $n_z=1$ is the $2p_0$ orbital.
Another possible orbital into which the $6h_{-5}$ wave function could evolve is the $2s$ state.
For the hydrogen atom or hydrogen-like ions in a magnetic field the $2p_{0}$ is stronger bound than the
$2s$ orbital. On the other hand, owing to the electron screening in multi-electron atoms in field-free space
the $2s$ orbital tends to be more tightly bound than the $2p_0$ orbital.
Thus, two states i.e. the $1s2p_02p_{-1}3d_{-2}4f_{-3}5g_{-4}$ state as well as the
$1s2s2p_{-1}3d_{-2}4f_{-3}5g_{-4}$ configuration are candidates for becoming the ground state
in the $S_z=-3$ set when we lower the field strength coming from the high field situation.

Analogous arguments lead to the three following
candidates for the ground state in case of the second crossover in the $S_z=-3$ subset which
takes place with decreasing field strength:
$1s2s2p_0 2p_{-1}3d_{-2}4f_{-3}$, $1s2p_0 2p_{-1}3d_{-1}3d_{-2}4f_{-3}$ and
$1s2s2p_{-1}3d_{-1}3d_{-2}4f_{-3}$.
It is evident that the one particle energies for the $3d_{-1}$ and $2p_0$ obey
$E_{3d_{-1}} > E_{2p_0}$ for all values of $\gamma$ since they possess the same nodal structure
with respect to the z-axis and only the $3d_{-1}$ possesses an additional node in the plane
perpendicular to the z-axis. For this reason the configuration
$1s2s2p_{-1}3d_{-1}3d_{-2}4f_{-3}$ can be excluded from our considerations of the ground state.
This conclusion is fully confirmed by our calculations.

Similar reasoning given in detail in ref. \cite{IvaSchm99} can
be repeated for two other ($S_z=-2$ and $S_z=-1$) subsets of states
and leads to the results presented in table \ref{tab:grcon}
and in Figure~\ref{foobar:c_f5}.

\begin{table}
\caption{The Hartree-Fock ground state configurations of the carbon atom
in external magnetic fields.
The configurations presented in the table are the ground state configurations
at $\gamma_{\rm min}\leq\gamma\leq\gamma_{\rm max}$.}
\begin{tabular}{llllrllllll}
no.&$\gamma_{\rm min}$&$\gamma_{\rm max_{}}$& The ground state configuration &$M$&$S_z$&$E(\gamma_{\rm min})$\\
\noalign{\hrule}
1&0     &0.1862   &$1s^2 2s^{2{}} 2p_0 2p_{-1}$             &$-1$ &$-1$&$-37.69096$\\
2&0.1862&0.4903   &$1s^22s2p_0 2p_{-1}2p_{+1}$              &$0 $ &$-2$&$-37.9334$\\
3&0.4903&4.207    &$1s^22s2p_0 2p_{-1}3d_{-2}$              &$-3$ &$-2$&$-38.3359$\\
4&4.207 &7.920    &$1s^22p_0 2p_{-1}3d_{-2}4f_{-3}$         &$-6$ &$-2$&$-41.7369$\\
5&7.920 &12.216   &$1s^22p_{-1}3d_{-2}4f_{-3}5g_{-4}$       &$-10$&$-2$&$-43.6397$\\
6&12.216&18.664   &$1s2p_02p_{-1}3d_{-2}4f_{-3}5g_{-4}$     &$-10$&$-3$&$-44.9341$\\
7&18.664&$\infty$ &$1s2p_{-1}3d_{-2}4f_{-3}5g_{-4}6h_{-5}$  &$-15$&$-3$&$-50.9257$\\
\end{tabular}
\label{tab:grcon}
\end{table}

\begin{figure}[b!] % fig 1
\centerline{\epsfig{file=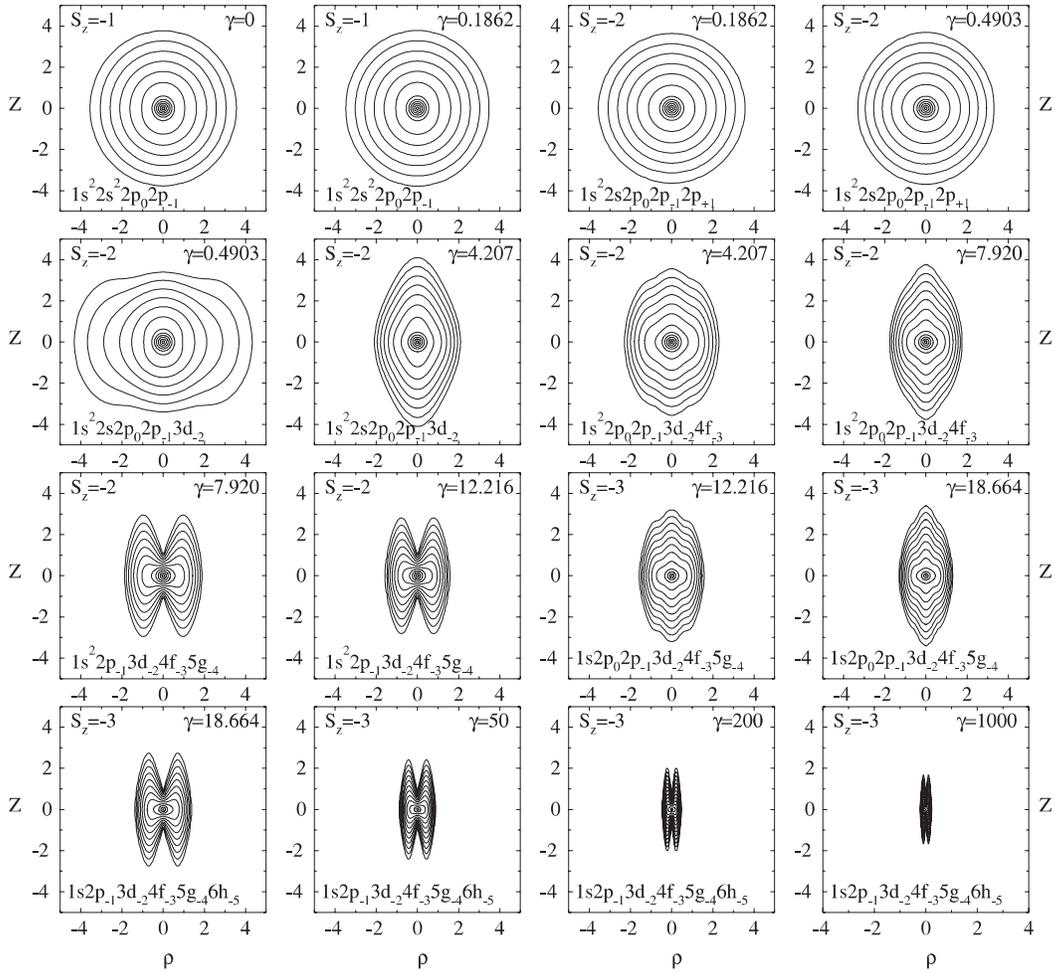,width=15cm}}
\vspace{10pt}
\caption{Contour plots of the total electronic
densities for the ground state of the carbon atom.
For neighbouring lines the densities are different by a factor of $e$.
The coordinates $z$, $\rho$ as well as the
corresponding field strengths are given in atomic units.}
\label{foobar:c_f60}
\end{figure}

Figure~\ref{foobar:c_f60} allows us to add some more informations 
to the considerations of the previous section.
This figure presents spatial distributions of the total electronic densities
for the ground state configurations of the carbon atom.
More precisely, it allows us to gain insights into the
geometry of the distribution of the electron density in space and in particular
its dependence on the magnetic quantum number and the total spin.
Thereby we can understand the corresponding impact on the total energy of the atom.
The first picture in this figure presents the distribution of
the electron density of the ground state of the carbon atom
at $\gamma=0$. The following pictures show the distributions of the electronic densities
at values of the field strength which mark the boundaries of the regimes of field
strengths belonging to the different ground state configurations.
For the high-field ground state we present the distribution of
the electronic density at the crossover field strength $\gamma=18.664$
and for three additional values of $\gamma$ up to $\gamma=1000$.

For each configuration the effect of the increasing field strength
consists in compressing the electronic distribution towards the $z$ axis.
However most of the crossovers of ground state configurations
involve the opposite effect which is due to the fact that they
are associated with an increase of the total magnetic quantum number
$M={\sum_{\mu=1}^6m_{\mu}}$.

For the lithium atom the analogous arguments and calculations \cite{IvaSchm98}
lead to the scheme presented in table \ref{tab:grli}.

\begin{table}
\caption{The Hartree-Fock ground state configurations of the lithium atom
in external magnetic fields.
The configurations, presented in the table are the ground state configurations
at $\gamma_{\rm min}\leq\gamma\leq\gamma_{\rm max}$.}
\begin{tabular}{llllrllllll}
no.&$\gamma_{\rm min}$&$\gamma_{\rm max}$& The ground state configuration&$M$&$S_z$&$E(\gamma_{\rm min})$\\
\noalign{\hrule}
1&0      &0.17633  &$1s^2 2s$                                &$0 $&$-1/2$&$-7.43275$\\
2&0.17633&2.1530   &$1s^22p_{-1}$                            &$-1$&$-1/2$&$-7.48162$\\
3&2.1530 &$\infty$ &$1s2p_{-1}3d_{-2}$                       &$-3$&$-3/2$&$-7.64785$\\
\end{tabular}
\label{tab:grli}
\end{table}

\section{Ground state electronic configurations in the
high-field regime}

Let us consider now the series of neutral atoms and positive ions
with $Z\leq 10$ in the high field domain which we define here
as the one,
where the ground state electronic configurations are fully spin polarised
(Fully Spin Polarised (FSP) regime $S_z=-N/2$).
The FSP regime supplies an additional advantage for
calculations performed in the Hartree-Fock approach, because
our one-determinant wave functions are
eigenfunctions of the total spin operator ${\bf S^2}$.
Starting from the high-field limit we investigate the electronic structure
and properties of the
ground states with decreasing field strength until we reach the first crossover
to a partially spin polarised (PSP) configuration with $S_z=-N/2+1$.

The approach to this investigation is very similar to that
described in the previous section and the picture of the
crossovers associated with the ground states of atoms
and positive ions $A^+$ is presented in table
\ref{tab:trans}.
It should be noted
that, for atoms with $Z\leq 6$  and ions with $Z\leq 7$,
the state $\left|1s^2\right>$ becomes
the ground state while lowering the spin polarisation
from the maximal absolute value $S_z=-N/2$ to $S_z=-N/2+1$.
For heavier atoms and ions we remark that the state $\left|1s^2\right>$
is not the energetically lowest one in the PSP subset
at magnetic field strengths for which
its energy becomes equal to the energy of the lowest FSP state.
For these atoms and ions the state $\left|1s^2 2p_0\right>$ 
is energetically lower
than $\left|1s^2\right>$ at these field strengths.
For atoms with $Z\geq 7$ and positive ions with $Z\geq 8$ the
intersection points between the state $\left|1s^2 2p_0\right>$ and
the energetically lowest state in the FSP subspace have to be calculated.
In result, the spin-flip crossover occurs at higher fields than this would be
in the case of $\left|1s^2\right>$ being the lowest state in the PSP subspace.
In particular, the spin-flip crossover for the neon atom is found to be
slightly higher than the point of the crossover $\left|2p_0\right>-\left|2p_0 3d_{-1}\right>$,
and, therefore, this atom has in the framework of the Hartree-Fock approximation
only two fully spin polarised configurations
likewise other neutral atoms and positive ions with $6\leq Z \leq 10$.
It should be noted that the situation with the neon atom can be regarded
as a transient one due to closeness of
the intersection $\left|2p_0\right>-\left|2p_0 3d_{-1}\right>$
to the intersection  $\left|2p_0\right>-\left|1s^2 2p_0\right>$.
This means that we can expect the configuration $\left|2p_0 3d_{-1}\right>$
to be the global ground state for the sodium atom ($Z=11$).
In addition an investigation of the neon atom carried out on
a more precise level than the Hartree-Fock method could also introduce some corrections
to the picture described above for this atom.

\begin{table}
\caption{Total energies (a.u.) of the neutral atoms and ions ${\rm A^+}$
at the crossover points of the ground state configurations.}
\begin{tabular}{lllllllll}
$Z$&$\gamma$&Atomic state(s)&$-E{\rm (Atomic)}$&Ionic state(s)&$-E{\rm (A^+)}$\\ \noalign{\hrule}
2&0.711&$\left|0_N\right>$, $\left|1s^2\right>$&2.76940&$\left|0_N\right>$&2.32488\\
\noalign{\hrule}
3&2.153&$\left|0_N\right>$, $\left|1s^2\right>$&7.64785&$\left|0_N\right>$&7.00057\\
&2.0718&$\left|1s^2\right>$&7.65600&$\left|0_N\right>$, $\left|1s^2\right>$&6.94440\\
\noalign{\hrule}
4&4.567&$\left|0_N\right>$, $\left|1s^2\right>$&15.9166&$\left|0_N\right>$&15.07309\\
&4.501&$\left|1s^2\right>$&15.91625&$\left|0_N\right>$, $\left|1s^2\right>$&15.01775\\
\noalign{\hrule}
5&8.0251&$\left|0_N\right>$, $\left|1s^2\right>$&28.18667&$\left|0_N\right>$&27.16436\\
& 7.957&$\left|1s^2\right>$&28.17996&$\left|0_N\right>$, $\left|1s^2\right>$&27.10004\\
\noalign{\hrule}
6&18.664&$\left|0_N\right>$, $\left|2p_0\right>$&50.9257&$\left|0_N\right>$&49.50893\\
&14.536&$\left|2p_0\right>$&47.23836&$\left|0_N\right>$, $\left|2p_0\right>$&45.77150\\
&12.351&$\left|2p_0\right>$&45.07386&$\left|2p_0\right>$, $\left|1s^2\right>$&43.72095\\
&12.216&$\left|2p_0\right>$, $\left|1s^2\right>$&44.9341&$\left|1s^2\right>$&43.70075\\
\noalign{\hrule}
7&36.849&$\left|0_N\right>$, $\left|2p_0\right>$&84.4186&$\left|0_N\right>$&82.58182\\
&30.509&$\left|2p_0\right>$&79.34493&$\left|0_N\right>$, $\left|2p_0\right>$&77.41246\\
&17.429&$\left|2p_0\right>$&66.72786&$\left|2p_0\right>$, $\left|1s^2\right>$&65.26170\\
&17.398&$\left|2p_0\right>$, $\left|1s^2 2p_0\right>$&66.69306&$\left|1s^2\right>$&65.25362\\
\noalign{\hrule}
8&64.720&$\left|0_N\right>$, $\left|2p_0\right>$&130.6806&$\left|0_N\right>$&128.4054\\
&55.747&$\left|2p_0\right>$&124.1125&$\left|0_N\right>$, $\left|2p_0\right>$&121.69825\\
&23.985&$\left|2p_0\right>$, $\left|1s^2 2p_0\right>$&94.3773&$\left|2p_0\right>$&92.78308\\
&23.849&$\left|1s^2 2p_0\right>$&94.3336&$\left|2p_0\right>$, $\left|1s^2 2p_0\right>$&92.62502\\
\noalign{\hrule}
9&104.650&$\left|0_N\right>$, $\left|2p_0\right>$&191.8770&$\left|0_N\right>$&189.1446\\
&92.624&$\left|2p_0\right>$&183.6944&$\left|0_N\right>$, $\left|2p_0\right>$&180.7819\\
&31.735&$\left|2p_0\right>$, $\left|1s^2 2p_0\right>$&128.1605&$\left|2p_0\right>$&126.4414\\
&31.612&$\left|1s^2 2p_0\right>$&128.1125&$\left|2p_0\right>$, $\left|1s^2 2p_0\right>$&126.2897\\
\noalign{\hrule}
10&159.138&$\left|0_N\right>$, $\left|2p_0\right>$&270.220&$\left|0_N\right>$&267.0112\\
&143.604&$\left|2p_0\right>$&260.2740&$\left|0_N\right>$, $\left|2p_0\right>$&256.8459\\
&40.672&$\left|2p_0\right>$, $\left|1s^2 2p_0\right>$&168.4734&$\left|2p_0\right>$&166.6327\\
&40.559&$\left|1s^2 2p_0\right>$&168.4217&$\left|2p_0\right>$, $\left|1s^2 2p_0\right>$&166.4863\\
\end{tabular}
\label{tab:trans}
\end{table}

Summarising our results we remark that 
the atoms and positive ions with $Z\leq 5$ have one FSP ground state configuration
$\left|0_N\right>$ whereas 
the atoms and ions with $6\leq Z\leq 10$ possess two such configurations 
$\left|0_N\right>$ and $\left|2p_0\right>$.

\begin{figure}[b!] % fig 1
\centerline{\epsfig{file=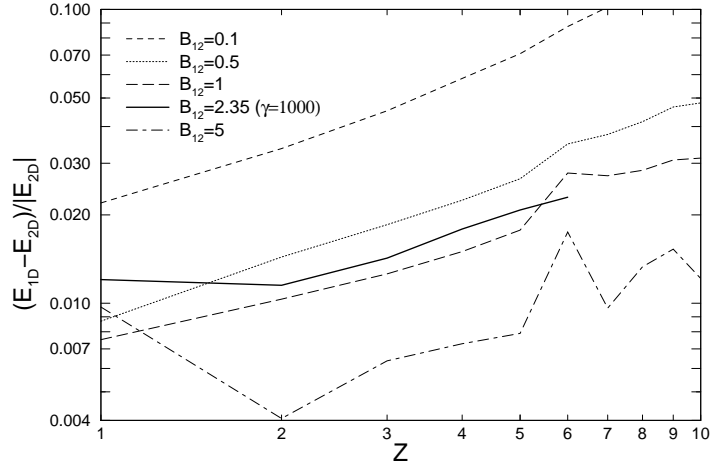,angle=270,width=3.5in}}
\vspace{10pt}
\caption{Relative errors of the total energy in the adiabatic approximation
depending on the charge of the nucleus. 
$E_{\rm 1D}$ -- the total energies in the adiabatic approximation 
($B_{12}=0.1,\ 0.5,\ 1,\ 5$ [\ref{three}], $B_{12}=2.35$ [\ref{two}]), 
$E_{\rm 2D}$ -- our two-dimensional mesh results. 
$B_{12}=B/10^{12}$G.}
\label{foobar:err}
\end{figure}

Possessing total energies for all the atoms with $Z\leq 10$ we can compare
these results with the adiabatic calculations \cite{Neuhauser,Godefroid}.
Both our results ($E_{\rm 2D}$) and calculations \cite{Neuhauser,Godefroid}
($E_{\rm 1D}$) are carried out in the adiabatic approximation.
The difference between $E_{\rm 1D}$ and $E_{\rm 2D}$
consists only in the usage of the adiabatic approximation for obtaining
energies $E_{\rm 1D}$ instead of exact solution of the Hartree-Fock
equations for $E_{\rm 2D}$.
Thus, the comparison of $E_{\rm 1D}$ and $E_{\rm 2D}$ allows us
to evaluate the precision of the adiabatic approximation itself
and obtain an idea of the degree of its applicability
for multi-electron atoms for different field strengths and nuclear charges.
All our values lie lower than the values of these adiabatic calculations.
It is well known, that the precision of the adiabatic approximation decreases with
decreasing field strength.
The increase of the relative errors with decreasing field strength is clearly
visible in the table.
On the other hand, the relative errors of the adiabatic approximation
possess the tendency to increase with growing $Z$, which is manifested
by the scaling transformation $E(Z,\gamma)=Z^2 E(1,\gamma/Z^2)$
(e.g. \cite{Rud94,Ivanov94}) well known for hydrogen-like ions.
The behaviour of the inner electrons is 
to some extent similar to the behaviour
of the electrons in the corresponding hydrogen-like ions.
Therefore their behaviour is to lowest order similar to the behaviour of the electron in
the hydrogen atom at magnetic field strength $\gamma/Z^2$ i.e. this
behaviour can be less accurately described by
the adiabatic approximation at large $Z$ values.
The absolute values of the errors in the total energy associated with the adiabatic approximation
are in many cases larger than the corresponding values of the ionisation energies.

\section{Mesh approach for
single-electron atomic and molecular systems
in strong electric fields}

In this section we present some features of our approach
With respect to its application to systems in strong
external electric fields and
correspondingly some relevant physical results.
In contrast to the situation discussed in the previous sections
atoms and molecules in external uniform electric fields
have no stationary states, because for every state there is
a probability that one or several electrons leave the system.
Thus, when switching on the external uniform electric field,
all the stationary states turn into resonances.
Using the complex form of the energy eigenvalues
\begin{eqnarray}
E=E_0-i\Gamma/2 \nonumber
\end{eqnarray}
one may consider quasi-stationary states of quantum systems similarly
to the stationary ones. In this approach the real part of the energy
$E_0$ is the centre of the band corresponding to the quasi-stationary
state and the imaginary part
$\Gamma/2$ is the half-width of the band which determines the lifetime of
the state.
In this communication we consider systems which can be described
by two-dimensional one-electron Hamiltonians.
These systems include the hydrogen atom and the ${\rm H_2^+}$
molecular ion in an electric field \cite{Ivanov94a,Ivanov98a,Ivanov2001}
and the hydrogen atom in parallel electric and magnetic fields
\cite{Ivanov2001,Ivanov83VLGU}.
The only electron present in such a system can leave it under
impact of the external electric field.
From the mathematical point of view the problem consists
in obtaining solutions of the
single-particle Schr\"odinger equation for this electron with the correct
asymptotic behaviour of the wavefunction as an outgoing
wave.
Currently we have three different possibilities for fixing
this asymptotics realised in our computational program:

\begin{figure}[b!]
\centerline{\epsfig{file=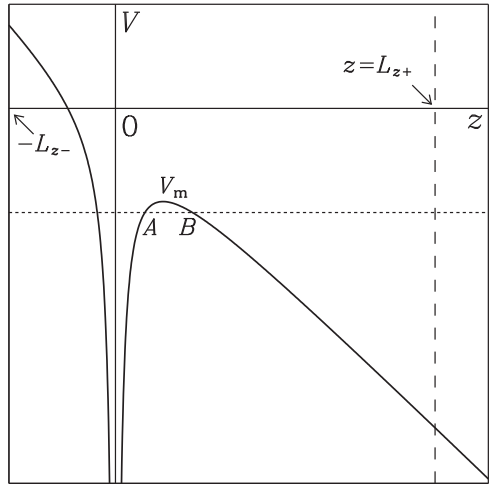,width=7cm}}
\vspace{10pt}
\caption{The potential energy for the hydrogen atom $V(\rho=0,z)$
in the external uniform electric field.}
\label{foobar:1r-fz}
\end{figure}

\begin{figure}[b!]
\centerline{\epsfig{file=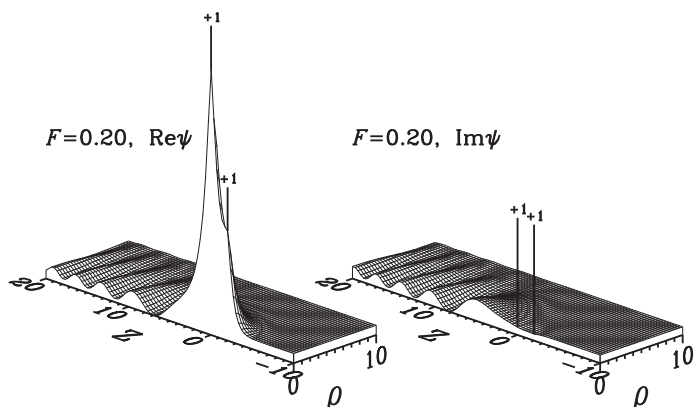,width=12cm}}
\vspace{10pt}
\caption{Real and imaginary parts
of the wavefunction of the $\rm H_{2}^+$
molecule in a longitudinal electric field $F$ (a.u.).}
\label{foobar:HHwave}
\end{figure}

{\it 1. Complex boundary condition method.}
This method is described in detail in ref. \cite{Ivanov98a}.
The method is based on the fact that the single-electron Schr\"odinger equation
for a finite system can be solved with the arbitrary precision
in a finite area both for stationary and for quasi-stationary eigenstates.
The case of stationary states is considered in \cite{ZhVychMat,Ivanov88}.
The approach for the quasi-stationary states will be discussed following \cite{Ivanov98a}.
Figure~\ref{foobar:1r-fz} presents the potential curve for the simplest
Hamiltonian of the hydrogen atom in an electric field
\begin{eqnarray}
H=-\frac 12\left(\frac{\partial ^2}{\partial \rho ^2}+
\frac 1\rho \frac \partial
{\partial \rho }+\frac{\partial ^2}{\partial z^2}-
\frac{m^2}{\rho ^2}\right)-\frac
1r-Fz
\label{HFSCHam}
\end{eqnarray}
$F$ is the electric
field strength multiplied by the charge of the electron.
Analogously to \cite{Ivanov83VLGU,Ivanov88}
the calculations can be carried out in an area
$\Omega $ which is finite along the direction $z$. 
For this coordinate we used
uniform meshes.  
The boundary of the area $z=-L_{z-}$ for
$z<0$ ($F\geq 0$) (Figure~\ref{foobar:1r-fz}) is determined from the
condition of small values of the wavefunction on the
boundary and, therefore, small perturbations introduced by
the corresponding boundary condition \cite{Ivanov88}.
The values of the wavefunction on the opposite boundary of
the area ($z=L_{z+}$) cannot be excluded from the
consideration. We consider non-stationary states of the
system decaying into continuum of free particles. In
this process an electron leaves the system in direction
$z\rightarrow +\infty $ and,
thus, an outgoing wave boundary condition is to be
established on $z=L_{z+}$. The form of this boundary
condition can be derived from the asymptotic behaviour of the
wavefunction for $z\rightarrow +\infty $ and
has the form
\begin{eqnarray}
\left. \frac{\partial \psi }{\partial z}+\left( \frac F{2k^2}-%
      i k\right) \psi\right| _{z=L_{z+}}=0\label{FeOutgo}
\end{eqnarray}
where $k=[2(E+Fz)]^{1/2}$ is the wavenumber.
Solving the Schr\"odinger equation with the Hamiltonian (\ref{HFSCHam})
and the boundary condition (\ref{FeOutgo}) established on a reasonable
distance $L_{z+}$ from the origin of the system we can obtain
the complex eigenvalues of the energy and the wavefunctions
of the type presented in the Figure~\ref{foobar:HHwave}.

This straightforward approach enables obtaining precise
results both for atoms and molecules from weak to moderate
strong fields (for instance for the ground state of the
hydrogen atom up to $F=0.20-0.25$~a.u.).

{\it 2. Classical complex rotation of the coordinate $z$ in the
form $z \rightarrow ze^{i\Theta}$.}
In this approach we have obtained precise results for atomic
systems in strong fields from the lower bound of the over-barrier regime
up to superstrong fields corresponding to regime $|{\rm Re}E|<<|{\rm Im}E|$
\cite{Ivanov2001}.
On the other hand, this method cannot be immediately applied
to molecular systems in our direct mesh approach \cite{Simon79,Ivanov2001}.

{\it 3. Exterior complex transformation of the coordinate $z$.}
In our numerical approach we transform the real coordinate $z$
into a curved path in the complex plane $z$.
This transformation leaves intact the Hamiltonian in the
internal part of the system, but supplies the complex rotation of
$z$ (and the possibility to use the zero asymptotic boundary conditions
for the wavefunction) in the external part of the system.
The transformation can be applied both for atoms and molecules and
provides precise results for fields from weak up to superstrong
with some decrease of the numerical precision in
the regime $|{\rm Re}E|<<|{\rm Im}E|$
\cite{Ivanov2001}.

The numerical results obtained by all three methods coincide
And are in agreement with numerous published data
on the hydrogen atom in electric fields (see e.g.
\cite{Kolosov87,Kolosov89,NicolaG,NicolaT} and references in \cite{Ivanov98a}).

Some of our results for
the hydrogen atom in parallel electric and magnetic fields \cite{Ivanov2001}
are shown in Table~\ref{tabHFB6}.

\begin{table}
\caption{The ground state of the
hydrogen atom in parallel electric and magnetic fields
at $F=0.1,\ 1,\ 5$.\label{tabHFB6}}
\footnotesize\rm
\begin{tabular}{@{}llllllllllll}
%\br
&\multicolumn{2}{c}{$F=0.1$}
&\multicolumn{2}{c}{$F=1$}
&\multicolumn{2}{c}{$F=5$}\cr
\cline{2-3}\cline{4-5}\cline{6-7}\\
$\gamma$&$E_0$&$\Gamma /2        $&$E_0$&$\Gamma /2         $&$E_0$&$\Gamma /2$\\
\noalign{\hrule}
$0   $&$-0.5274183$&$7.26904(-3)$&$-0.6243366$&$0.6468208 $&$-0.1350071$&$3.083929   $\\
$0.01$&$-0.532390 $&$7.2624(-3) $&$-0.629329 $&$0.646812  $&$-0.140005 $&$3.083925   $\\
$0.1 $&$-0.574600 $&$6.6392(-3) $&$-0.673584 $&$0.646053  $&$-0.184739 $&$3.083589   $\\
$1   $&$-0.8443098$&$9.5923(-5) $&$-1.0421379$&$0.577291  $&$-0.6077008$&$3.050207   $\\
$10  $&$-1.7498730$&---          &$-1.9579187$&$0.1173924 $&$-2.375552 $&$1.955678   $\\
$100 $&$-3.790110 $&---          &$-3.8219215$&$4.9988(-5)$&$-4.3806709$&$0.4909467  $\\
$1000$&$-7.66247  $&---          &$-7.66807  $&---         &$-7.82561  $&$1.04701(-2)$\\
%\br
\end{tabular}
\end{table}

\begin{figure}[b!] % fig 1
\centerline{\epsfig{file=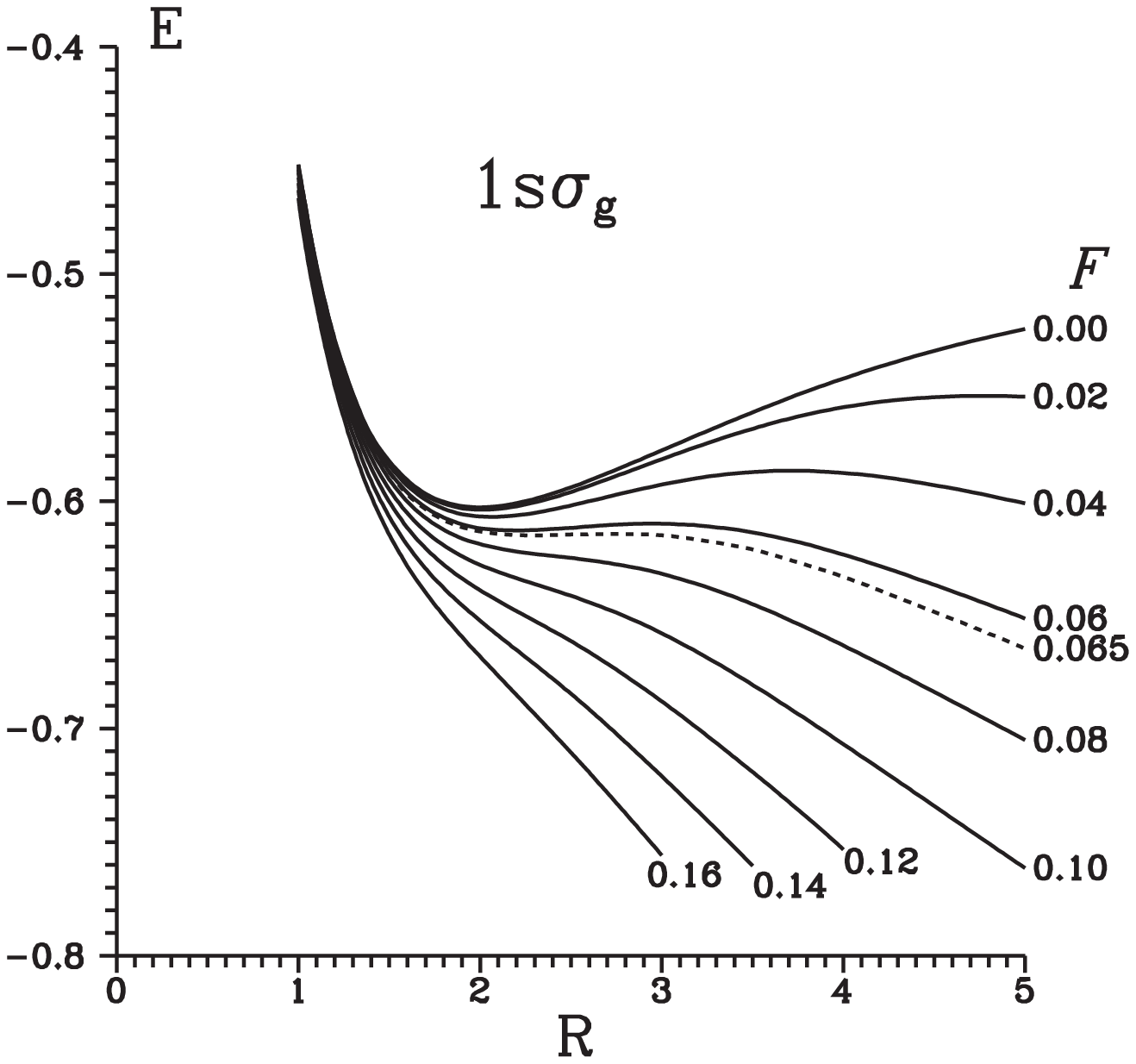,width=7cm}\hspace*{0.8cm}\epsfig{file=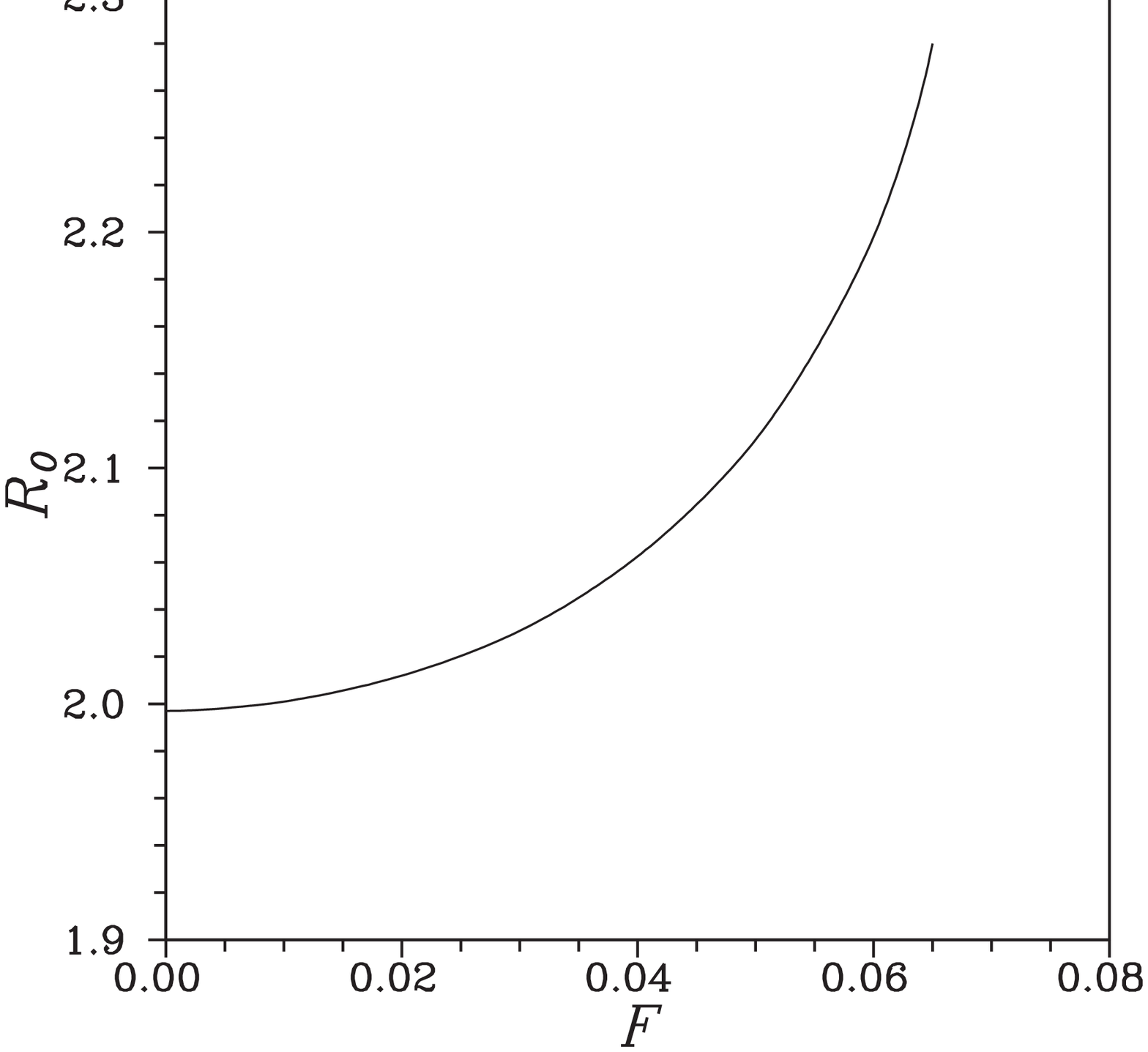,width=4.7cm}}
\vspace{10pt}
\caption{Left -- Potential curves for the ground state of the $\rm H_{2}^+$
molecule in longitudinal electric field $F$ (a.u.).
Right - The equilibrium internuclear distance of the molecule
$\rm H_{2}^+$ as a function of the applied electric field strength.}
\label{foobar:HHPEFGRA}
\end{figure}

\begin{table}
\caption{Equilibrium internuclear distances and corresponding energies
and half-widths of the energy of the ground state of
$\rm H_{2}^+$ molecule in
longitudinal electric fields. \label{tabthree}}
\footnotesize\rm
\begin{tabular}{@{}llllllllll}
%\br
$F$&$R_0$&$E_0$&$\Gamma /2$
&$F$&$R_0$&$E_0$&$\Gamma /2$
\cr
\noalign{\hrule}
0.00 &1.997&-0.60264&---      &0.04 &2.062&-0.60686&1.23(-19)\cr
0.01 &2.001&-0.60289&---      &0.05 &2.112&-0.60943&7.32(-15)\cr
0.02 &2.012&-0.60366&---      &0.06 &2.198&-0.61285&1.49(-11)\cr
0.03 &2.031&-0.60497&---      &0.065&2.28 &-0.61501&3.94(-10)\cr
%\br
\end{tabular}
\end{table}

The second system which we present in this section is the
hydrogen molecular ion $\rm H_{2}^+$ in a strong longitudinal electric field.
Our approach allowed us to carry out the first correct
consideration of this system \cite{Ivanov94a} and,
in particular, to obtain the potential curves of its
ground state, presented in Figure~\ref{foobar:HHPEFGRA} (left).
The minima in these curves give the equilibrium
internuclear distances, presented in this Figure (right)
as a function of the electric field strength.
One can see that at a critical value of the electric field
about 0.065~a.u. the minimum disappears and, thus,
above this critical value the hydrogen molecular ion cannot exist.
This critical value of the maximal electric field
$F_{\rm c}=0.065{\rm a.u.}=3.3{\rm V/\AA}$ for the
molecule $\rm H_{2}^+$ is in
a good agreement with experimental results by \cite{Bucksbaum90}.
According this work $\rm H_{2}^+$ molecule may exist in laser
beam fields with intensity less than $10^{14}{\rm W/cm^2}$ which
corresponds to $3{\rm V/\AA}$, and does not exist in more intense
fields.

\section{Conclusions}

In this communication we have presented a 2D fully numerical
mesh solution method in its various applications to atoms
and simple diatomic molecules in strong external electric
and magnetic fields.
Specifically these are calculations of atoms with $Z\leq 10$ and their
positive ions in strong magnetic fields and the comprehensive
investigation of the electronic structure of the ground states
of the Li and C atoms in arbitrary magnetic fields.
For the carbon atom seven different
electronic ground state configurations for
different domains of the magnetic field strength 
have been found.
The investigation of the series of atoms with $Z\leq 10$
in very strong magnetic fields enables us to evaluate
the applicability of the adiabatic approximation and
to show its decreasing precision for heavier atoms.

The mathematical technique developed for solving
Schr\"odinger equations for quasi-steady states
allowed us to obtain a series of results for the hydrogen
atom in parallel electric and magnetic fields and
for the ${\rm H_2^+}$ ion in strong electric fields.

Thus, the method described above allows us to obtain
a number of new physical results partially presented
in this communication.
These calculations are carried out in the Hartree-Fock
approximation for multi-electron systems and are
exact solutions of the Schr\"odinger equation for
the single-electron case.
As the following development of the method we plan to implement 
the configuration interaction approach in order 
to study correlation effects in
multi-electron systems both in electric and magnetic fields.

{}

\end{document}